**Research in context**

**Evidence before this study**

The global disruption of life by COVID-19 and the persistent uncertainties related to widespread mobility restrictions has spurred multiple questionnaire studies on self-perceived anxiety and sleep. The prevalent concern is about disrupted sleep and understandably the impact has been greatest on front-line health care workers. Less is known about post-lockdown sleep in the general working population whose wellbeing will be critical for restarting damaged economies. Widespread lockdowns also affect physical activity, another powerful influencer of wellbeing. There has been virtually no research into collective sleep and physical activity shifts in the working public.

**Added value of this study**

We analyzed longitudinal sleep/activity wearable tracker data from ~1800 office workers collected before the COVID-19 outbreak, through incremental movement restrictions culminating in lockdown that permitted socially distanced walking, running and cycling. In addition to characterizing objective measures of sleep and physical activity across the entire sample, we demonstrate how heterogenous groups with different sociodemographic characteristics are affected by using novel rest activity rhythm and hierarchical clustering approaches.

**Implications of the available evidence**

Contrary to popular expectation, sleep shifted later but may otherwise benefit from lockdown. Substantial drop in physical activity is of greater concern. Longitudinal monitoring of rest-activity rhythms through incrementally severe movement restrictions revealed heterogenous patterns of sleep and physical activity across a population. Wide-scale adoption of the methods described here may help identify groups at risk should further or extended lockdowns occur.

# COVID-19 Related Mobility Reduction: Heterogenous Effects on Sleep and Physical Activity Rhythms

J.L. Ong PhD[1], T.Y. Lau BA[1], S.A.A. Massar PhD[1], Z.T. Chong BSc[2], B.K.L. Ng PhD[2], D. Koek MSc[2], W. Zhao PhD[2,3], B.T.T. Yeo PhD[1,4,5], K. Cheong PhD[2] and M.W.L. Chee MBBS[1]*

[1]Centre for Sleep and Cognition, Yong Loo Lin School of Medicine, National University of Singapore, 12 Science Drive 2, Singapore 117549, Singapore.
[2]Health Promotion Board, 3 Second Hospital Ave, Singapore 168937, Singapore.
[3]Centre for Quantitative Medicine, Duke-NUS Medical School, 2 College Rd, Singapore 169857, Singapore.
[4]Department of Electrical and Computer Engineering, 4 Engineering Drive 3, National University of Singapore, Singapore 117583, Singapore.
[5]N.1 Institute for Health, National University of Singapore, 28 Medical Drive, Singapore 117456, Singapore.

* Corresponding author:
Michael W.L. Chee
Professor and Director,
Center for Sleep and Cognition,
Yong Loo Lin School of Medicine,
12 Science Drive 2,
National University of Singapore,
Singapore 117549.
Email: michael.chee@nus.edu.sg

## ABSTRACT

**Background**

Mobility restrictions imposed to suppress coronavirus transmission can alter physical activity (PA) and sleep patterns. Characterization of response heterogeneity and their underlying reasons may assist in tailoring customized interventions.

**Methods**

We obtained wearable data covering baseline, incremental movement restriction and lockdown periods from 1824 city-dwelling, working adults aged 21-40 years, incorporating 206,381 nights of sleep and 334,038 days of PA. Four distinct rest activity rhythms (RAR) were identified using k-means clustering of participants' temporally distributed step counts. Hierarchical clustering of the proportion of time spent in each of these RAR revealed 4 groups who expressed different mixtures of RAR profiles before and during the lockdown.

**Findings**

Substantial but asymmetric delays in bedtime and waketime resulted in a 24 min increase in weekday sleep duration with no loss in sleep efficiency. Resting heart rate declined ~2 bpm. PA dropped an average of 38%. 4 groups with different compositions of RAR profiles were found. Three were better able to maintain PA and weekday/weekend differentiation during lockdown. The least active group comprising ~51% of the sample, were younger and predominantly singles. Habitually less active already, this group showed the greatest reduction in PA during lockdown with little weekday/weekend differences.

**Interpretation**

Among different mobility restrictions, removal of habitual social cues by lockdown, had the largest effect on PA and sleep. Sleep and resting heart rate unexpectedly improved. RAR evaluation uncovered heterogeneity of responses to lockdown and can identify characteristics of persons at risk of decline in health and wellbeing.

**INTRODUCTION**

Adequate sleep and physical activity (PA) are two of the triad of lifestyle factors critical to multiple aspects of health and wellbeing. Mobility restrictions imposed to contain spread of COVID-19 massively disrupted the daily routines of people worldwide, but particularly the structured lives of city-dwellers who represent 55% of the world's population[1]. Lockdowns interfere with time cues that mark when to wake-up, commute to work, organize meals, care for dependents, socialize, recreate and wind down to sleep[2]. Heterogeneity in changes to sleep and PA in response to lockdown can increase health disparities and are the focus of this study.

Urban life predisposes persons to late bedtimes on weekdays and greater sleep extension on weekends[3] which elevates risk for metabolic dysfunction[4]. During the pandemic, greater anxiety[2] and increased engagement on electronic media[5] for social engagement and news gathering could drive bedtimes later for some, curtailing sleep. In contrast, working from home[6] could afford greater flexibility in scheduling, avoidance of commuting and reduced stress for others, which could facilitate sleep. Pre-print findings in participants of a disease-monitoring consortium suggest the latter[7].

Physical activity benefits musculoskeletal, cognitive[8], cardiometabolic health[9] and sleep[10]. Additionally, exercise outdoors positively influences mental wellbeing[11] and morning light exposure serves to synchronize the circadian clock[12], checking the innate tendency to sleep progressively later. Even if one does not have time for intentional exercise, walking on the way to and from work can contribute significantly to physical activity[13,14]. Mobility restrictions serve to severely limit commuting but the extent to which outdoor activities are limited[15] varies significantly across countries[7] modulating this limitation.

In addition to duration of sleep and physical activity of themselves, their respective timing and distribution matter. For example, it may be preferable to accumulate physical activity across the day rather than to concentrate it within a short period while retaining extended sedentary time[16]. How incremental mobility restriction upsets the rhythm of sleep and daytime physical activity is therefore of interest.

Since August 2018, over a thousand young working adults in Singapore were provided with a Fitbit™ Ionic wearable sleep and activity tracker to evaluate health behavior. The intention was to examine the effects of lifestyle factors and their modulation on chronic non-communicable disease risk. This ongoing study provides a unique opportunity to characterize how COVID-19 associated movement restrictions shift sleep and physical activity patterns from previously established baselines using objective, longitudinal measurement as distinct from questionnaire-based studies[5,17]. In addition to whole sample characterization, we used machine-learning data-clustering to show how lockdown results in heterogenous sleep and physical activity transformations in different sociodemographic groups. These differentiate our approach from data mining using public tools[18] to access aggregated data that offers broader coverage but is less useful for formulating targeted interventions[7].

## RESULTS

**Key characteristics of the sample and time series data**

The data reported here represents 206,381 nights of sleep and 334,038 days of physical activity data collected from 1824 persons from Jan 3 – Apr 29th 2019, and Jan 2 – Apr 27th 2020. These dates refer to the morning of each sleep/daytime activity record, such that sleep always preceded activity for that date. Participants were between 21-40 years of age, 51.64% of whom were women. Most were office workers who commuted to Singapore's Central Business District. They were relatively well educated (86.1% college degree) and earned a median salary of SGD 4000-5999. Singles comprised 57.8%, married persons with children 21.9%. Other details about sociodemographic features can be found in Table 1. Average weekday daily commute time reported in this sample was 1.92h (SD: 0.71h), so time taken to travel to and fro work could be estimated to be ~1h each way. Details on the requirements for participation are described in Materials and Methods.

Singapore responded to the COVID-19 pandemic by instituting a host of infection control measures that ramped up from the announcement of the first case in Singapore (Jan 23), to the raising of its health alert (Feb 7), followed by stay-at-home orders for overseas returnees, announcement of border closures (Mar 16) and closing of pubs (Mar 26) constituting the first phase of movement restrictions (Fig. 1). An outbreak of infections in construction worker dormitories triggered a lockdown that caused cessation of non-essential services (Apr 7). Schools, malls and retail outlets were closed, and workers had to work from home. Outdoor activities were confined to walking, running and cycling with appropriate social distancing. This can be considered the second, 'lockdown' phase of mobility restriction (Fig. S1).

**Baseline sleep and physical activity characteristics across the sample**

Consistent with the highly structured nature of city life, participants kept to a regular sleep and wake schedule during both weekdays and weekends. Sleep timing and duration was highly consistent from week to week, as well as across two equivalent months in January 2019 and 2020 (Table 2; Fig. 1 A-D). On average, at baseline, bedtime was 00:13, waketime was 07:07 and total time in bed was 6.9 h (total sleep time: 6.0h) on weekdays. On weekends, bedtime was 00:53, waketime was 08:26 and total sleep duration was 7.5 h (total sleep time: 6.5h). Weekday-weekend sleep extension of 38 minutes during this period was present suggesting inadequacy of habitual pre-pandemic weekday sleep. Sleep efficiency, defined as percentage of time spent actually asleep out of total time in bed, and which is often considered as an objective marker of sleep quality, was generally high at ~87% (Fig. 1E).

Physical activity on weekdays showed a creep upwards from Monday to Friday, peaking on Saturdays and sharply dropping on Sundays before starting another cycle (Fig. 1F). Overall, participants' step counts hovered close to the popularized '10 000 steps a day' threshold week on week, averaging about 9700 steps. Of this, between 35-50 minutes was in the form of moderate-to-vigorous physical activity (Fig. 1G).

**Sleep Changes with Incremental Mobility Restrictions**

Sleep patterns were neither affected by the raising of the national health alert level, nor by the WHO pandemic announcement. Only subsequent to border closures and institution of non-mandatory split-teams at work, did weekday sleep timings start to shift (Fig. 1 A-D). Weekday bedtimes were delayed by 11 min and wake times by 22 min. The imposition of lockdown brought on a further shift in weekday bedtimes by about 19 min (total shift of 30 min from baseline) and a delay in wake times by 33 min (total of 55 min from baseline).

Weekend shifts in bedtime were more modest, with no significant shift until the lockdown period and thereafter, a total delay of 19 minutes from baseline. Despite the magnitude of this shift being more modest than the weekday shift, the most socio-economically impactful public holiday period in Singapore, Chinese New Year, only shifted 'weekend' sleep timing by less 12 minutes. As weekend waketimes were also delayed by 31 minutes there was increased total weekend sleep duration.

Overall, total weekday sleep duration increased by about 24 minutes during lockdown compared to the habitual weekday baseline; the increase in sleep duration over weekends was smaller 13 minutes. The shifts in sleep timing over weekdays and weekends resulted in the attenuation of the habitual and robust weekend sleep extension to 26 min during lockdown. Sleep efficiency remained at ~87% throughout the mobility restriction.

**Physical Activity Changes with Incremental Mobility Restrictions**

Similar to the pattern of changes observed with sleep, step counts started to decline only after border and pub closures, dropping on average by ~15% steps from baseline. To put this decrement in perspective, it was larger in percentage and absolute count differential elicited by rewarding increased physical activity during a previous physical activity intervention program[19]. Following the imposition of the lockdown, there was a further overnight decline, resulting in a cumulative 38% decline in average step counts. Before the pandemic, participants were relatively more active on weekdays. This pattern reversed following the lockdown.

Changes in moderate-to-vigorous physical activity (MVPA) broadly occurred in synchrony with step counts, although decreasing less, about 31% over the week on average. However, there were important differences. As MVPA likely incorporates intentional physical exercise, it would likely be part of entrenched behavior that would be more robust to change, even with lockdowns. This can be seen by continuation of baseline levels of MVPA until pubs and gyms were closed where upon a downtrend commenced. Lockdown elicited a second decrement in MVPA possibly contributed by stadium closures.

**Systematic patterns of heterogeneity in sleep and physical activity uncovered with RAR analysis**

In order to investigate concurrent changes to the magnitude of physical activity, as well as the duration and timing of sleep and activity within an individual across days, we explored changes to 24h rest-activity rhythm (RAR) profiles with increasing mobility restrictions. Characterization of patterns and shifts in 24h sleep-wake behavior based on actigraphic monitoring of locomotor activity has been an area of interest for circadian researchers for many years, as disruptions to the timing, amplitude and regularity of these RARs (e.g. by ageing or disease) can influence health and behaviour[20,21]. While traditional methods typically fit a 24h sinusoid to describe these rhythms, we have shown that this is often a poor representation of real-world rhythms in populations with relatively fixed weekday schedules. We employed a data-driven approach to identify profile clusters, or 'basis patterns' of these RARs, using k-means clustering of intraday log-transformed step counts (natural log) in 15-min bins across the day. For this analysis, we used all valid days with a minimum daily wear time of 13h from all participants in Jan-Apr 2020 (125,851 days).

We identified 4 distinct RAR profiles: 'Active 3-Peak Early', '3-Peak Middle', 'Active 2-Peak Later' and 'Inactive 3-Peak', describing both magnitude as well as timing of preferred daytime activity (Fig. 2A). Before the lockdown, weekdays tended to consist of more 'Active 3-Peak Early' and '3-Peak Middle' days (Fig. 2 B), indicating the strong influence of work as a frame around which life is organized (Peak 1: Travelling to work, Peak 2: Travelling to lunch, Peak 3: Travelling home). In contrast, weekends tended to consist of more 'Active 2-Peak Later' and 'Inactive 3-Peak' days, indicating temporally less structured / lower magnitude of daytime activity. During the lockdown, both 'weekend' patterns increased their expression during weekdays.

A control analyses performed on Jan 2019, Jan 2020 and during the lockdown separately revealed that before the restrictions were imposed, the clustered profiles were virtually identical (r > 0.99; Fig. S2, A and B). During the lockdown period, profiles became attenuated but were still highly similar to the profiles estimated using the full dataset (r: 0.84-0.98; Fig. S2C).

**Impact of lockdown on rest activity rhythms differs dramatically across individuals**

To identify groups of individuals who show similar changes in RAR profile composition with increasing mobility restrictions, we performed hierarchical clustering on the proportion of time each individual spent in the four canonical RAR profiles before and during the Apr 7 lockdown (Fig. 3A), and compared sleep and physical activity changes during the lockdown to a baseline period (Table 3, Fig. S3 to S4). This analysis was confined to 670 individuals who had at least 60% of data before and during the lockdown (missing days were excluded from the proportions calculation, such that proportion of time spent summed to 1 both before and during the lockdown). Supplementary analysis revealed that compared to excluded participants, included participants tended to be older (~1y) and had a higher proportion of individuals married with children (27% vs. 19%).

Group 1, whose dominant profile alternated between "3-Peak Middle" and "Active 2-Peak Later" before the lockdown showed the most drastic change to a predominant "Inactive 3-Peak" during the lockdown (>70% of the time; Fig. 3B-C), with a clear abolition of weekday-weekend differences in RAR profiles. They were the single largest group comprising about 51% of the sample who were the least physically active compared to the other 3 groups at baseline (9517 steps in Group 1 compared to 12011, 10960, 11275 steps in Groups 2, 3 and 4 respectively; all $p$s < .001). This group also showed the largest drop in step count - a massive 51% reduction in steps relative to baseline (-4833 steps in Group 1 compared to -2742, -2098, and -2966 in Groups 2, 3 and 4 respectively; all $p$s < .001). This group also slept and woke late. Total sleep time was not adversely affected by the lockdown, and in fact, slightly increased by 17.5 min ($p$ < .001). Interestingly, in contrast to their strongly attenuated weekday /weekend activity difference (pre-lockdown attenuation = -590 steps, post-lockdown attenuation = +289 steps, p < .001), there was only relatively, modest attenuation of weekday/ weekend difference in sleep duration (pre-lockdown difference +37 min TST, during-lockdown, +23 min difference in TST, p < .001). This group was over-represented by younger, single persons who likely have more flexibility with day-day schedules.

Group 2 comprised 14% of persons whose dominant RAR was 'Active 3-Peak Early' (50-70% of the time; Fig. 3B-C) before and during the lockdown, although there was a slight reduction in proportion of the dominant profile on weekdays during the lockdown. This group was the most physically active, averaging around 12011 steps before the pandemic and 9268 steps when the lockdown was imposed. This group was over-represented by married couples with children. Consistent with having childcare responsibilities that would require them to wake earlier in the day, this group tended to retain their habitual sleep / wake timings as well as sleep duration. It is possible that the slight reduction in their dominant profile on weekdays during the lockdown suggests that they did not have to wake up as early to prepare themselves for work and their children for school.

Group 3 expressed a clear weekday dominant '3-Peak Middle' RAR profile and weekend dominant 'Active 2-Peak Later' RAR profile, which was preserved during the lockdown (Fig. 3B-C). This group had moderate step counts (between Group 1 and 2) and moderate sleep-wake timings. Apart from a later sleep and wake timing, they bore significant similarities to Group 2 in terms of their consistency and intensity of physical activity. This group averaged 10960 steps before lockdown and dropped the least of all 4 groups and remarkably, maintained their habitual duration of MVPA even through lockdown ($p = .50$). This group was best able to maintain weekday weekend differences in routines. Group 3 was strongly dominated by persons with a college degree but had even representation in terms of family status.

Group 4 was mainly characterized by a dominant "Active 2-Peak Later" RAR profile on both weekdays and weekends (Fig. 3B-C), that was only slightly attenuated during the lockdown as proportion of time spent in the dominant state increased on weekdays. It was the second largest group in the sample (27%) resembling Group 1 in terms of late sleep and wake timings ($p = .99$, $p = .22$ respectively) but its members were more physically active ($p < .001$), resembling Group 3 in consistency of steps, and intensity of overall physical activity ($p = .48$, $p = .49$ respectively). This group also had more individuals who were married with children.

**DISCUSSION**

In young adult office workers living in a densely populated city, incremental mobility restrictions to stem the spread of COVID-19 showed the greatest impact on physical activity. A full lockdown, involving cessation of regular office work, school closures and limitation on interpersonal interaction to nuclear family members, but not lesser restrictions had the clearest impact. Rest activity rhythm analysis revealed significant heterogeneity the ability to maintain habitual routines despite this major societal upheaval. Against popular notions that sleep could be significantly disrupted, we found only modest evidence for this in the early period following lockdown.

The strong influence of going to the office on physical activity accrues from three observations. Most compelling was the overnight, 23% drop in step count when offices were closed but walking, running or cycling outdoors with social distancing, were allowed. Second, post-lockdown, weekend steps exceeded weekday steps, reversing the baseline pattern likely driven by commuting-related weekday steps. Thirdly, the ~50-minute delay in wake time and later commencement of a diminished morning bump in physical activity.

The widespread availability of broadband internet, mobile video conferencing and establishment of business continuity plans have made it more feasible for persons to work from home than ever before. However, until lockdown, physical activity records suggest that the status quo of going to work largely continued. Alternative arrangements such as working from home[22] or shortening the workweek[23] have been extensively debated [6] with relatively little society-wide change. The extensive lockdowns globally, provide an unprecedented opportunity to critically reappraise the pragmatics of these issues.

Contrary to widespread expectation that the pandemic would elevate anxiety which would be reflected in poorer sleep[2,17], we did not find support for this. Total time in bed *increased* on weekdays and weekend sleep extension, a sign of inadequate weekday sleep, decreased. Sleep efficiency was maintained from baseline through lockdown suggesting no significant disruption in sleep initiation or continuity. Perhaps more interestingly, resting heart rate, an indicator of cardiovascular risk[24], dropped by ~2 bpm.

For persons in Groups 1 and 4, not having to wake up early to go to the office, likely surfaced innate delayed sleep timing, thus lessening the impact of keeping to traditional work hours. The overall later timings of bedtime and waketime, even for earlier sleepers and risers relative to results published elsewhere[25], could be indicative of Singapore's 'westerly' position relative to its assigned time zone. Later bedtimes and shorter nocturnal sleep co-occur with chronic, longitude-related exposure to later evening light[26].

The robustness of RAR profiles revealed over comparable weeks in January 2019 and 2020 speaks to their utility as basis patterns whose automated detection reduces the dimensionality of longitudinal, sleep and physical activity data. This could simplify the identification of persons who might benefit from customized counsel during extended lockdowns. When data on the long-term impact on health, wellbeing or productivity measurements emerge, RAR may be refined as a public health predictive tool.

Averaged over weeks or months, a step-count close to 10,000, preservation of MVPA, a sleep duration closer to that recommended, and a lower resting heart rate are desirable. Greater sleep regularity[27,28] and lesser weekday-weekend differences[29] in sleep timing are considered favorable to health. However, the attenuation in weekday/weekend physical activity differences seen here likely reflects social isolation[30] that has long term consequences on mental wellbeing.

Given these considerations, Group 3 may have the best combination of outcome variables, being associated with the best preserved MVPA and well-preserved weekday-weekend variation in physical activity together with more consistent bedtimes and waketimes. In contrast, over half the sample in Group 1 showed a ~51% reduction in step counts and relatively lesser reduction in weekend sleep extension in response to mobility restriction. Although not statistically significant, there was a trend for resting heart rate, a proxy for cardiovascular risk, to diverge between these two groups such that over a longer period of observation, Group 1 might be at higher risk.

Sociodemographic variables influence RAR pattern shifts. Group 1 was disproportionately represented by singles who may be working even longer hours from home, resulting in less time for physical activity and profound loss of differentiation between weekends and weekdays. This group of individuals could benefit from setting regular routines to provision time for rest, physical activity and work, and to sleep and wake at fixed times at a time when social cues are disrupted. In contrast, Group 2 who wake up the earliest and also sleep earlier are over-represented by persons who are married with children. Having the responsibility to care for the latter would serve to drive more temporally structured activity, leading to the preservation of a three peaked RAR. The repetition of this routine daily without recourse to weekend family outings, on the other hand serves to obscure weekday/weekend activity differences.

**Limitations**

Limitations include a relatively limited time window of observation, the likely exclusion of persons with highly disturbed sleep schedules who might have stopped contributing their data, and the non-availability of health biomarkers. The collection of subjective sleep quality measures might also have revealed changes not uncovered by objective sleep markers. For example, increased hypothalamic-pituitary-adrenal (HPA) axis reactivity that relates to perceived stress, seems to relate more to sleep quality more than sleep duration[31]. Related to this, interpretation of the lowered resting heart rate findings could have benefited from information about participants' self-perceived stress. As most of the participants have existing jobs, which for the present time appear to be preserved by various governmental financial supports, the sleep patterns here are likely, less affected than for a socio-economically distressed sample.  Finally, as further lockdowns are likely and the period of exposure to widespread mobility restrictions may be extended, the findings herein may be time bounded in their generalizability.

**Conclusion**

Longitudinal monitoring of rest-activity rhythms through incrementally severe movement restrictions during the COVID-19 pandemic revealed heterogenous patterns of sleep and physical activity. Widescale adoption of the methods described here may help identify persons or groups at risk should further lockdowns occur as well as to gauge the impact of different lockdown unwinding policies.

**MATERIALS AND METHODS**

**Data source**

Data was obtained from the 'Health Insights Singapore' (hiSG) study, a longitudinal population-health study by the Health Promotion Board using wrist-worn wearable technology. Initiated in August 2018, the study recruited 1,951 young adults working in the Central Business District aged 21-40 years. Participants were given devices (Fitbit™ Ionic, Fitbit™ Inc, San Francisco, CA) to track their activity/sleep and installed a mobile application to complete surveys over a period of 2 years. Participants were rewarded with points convertible to vouchers if they wore the tracker daily, logged sleep, meals, and completed surveys and were allowed to keep the device conditional on meeting study requirements. Demographic, health and lifestyle questionnaires were administered at study commencement. A second survey conducted in February 2020 was used to update any changes to family status. The National Healthcare Group Domain Specific Review Board approved the study protocol. Informed consent was obtained from all subjects prior to study participation.

To evaluate the impact of the COVID-19 pandemic, we studied data gathered between 2 Jan 2020 and 27 Apr 2020, starting three weeks before the first case was reported in Singapore ('Baseline', Jan 2-22) and ending three weeks into the lockdown enforced by the Singapore government ('Lockdown', Apr 7-27). To compare this to an equivalent period in 2019, we used data from 3 Jan – 29 Apr 2019. Only individuals who had valid data on both years after filtering (see below) were included. This comprised 206,381 nights of sleep and 334,038 days of physical activity from a final sample of 1824 individuals. All dates presented refer to the morning of each sleep/daytime activity record, such that sleep always preceded activity for that date.

**Tracker-based data**

Sleep and activity data for each participant were extracted from the Fitbit™ API. The activity data comprised daily total steps, moderate-to-vigorous activity (MVPA) minutes (sum of fairly and very active minutes), resting heart rate levels and intraday step counts in 15 min intervals. For the comparison of measurements across time, data was filtered to remove days when participants did not wear the Fitbit™ for at least 8h/day or when atypical activity levels were observed. This was defined as records with 1) total daily steps > 50000, 2) total daily steps > 40000 and sedentary minutes > 1320 min, 3) sedentary minutes = 1440 min, and 4) no resting heart rate. Between 1041 to 1562 (mean: 1375) participants contributed to each timepoint for physical activity data. Wear time averaged 18h to 19h (mean: 18.5h) for each day. Not everyone contributed data on all the dates, but the large sample made it less likely that any individual's data would significantly alter group means.

For the computation of RAR profile clusters, only days containing intraday steps with no indicators of long periods of non-wear, as well as days with at least 13h of total wear time were used, for a final sample of 125,851 valid days.

Sleep data consisted of bedtimes, wake times, time in bed (TIB), total sleep time (TST), and time spent awake after sleep onset (WASO). Sleep efficiency was (100*TST/TIB). As in our prior work, we limited analyses to only nights with heart rate-derived sleep staged data, as this ensured proper wear time during the night and excluded records manually adjusted by the user. Records that indicated <4h TIB or >12h TIB were also excluded from the calculation of sleep variables, as it could indicate inappropriate detection of sleep by the algorithm (e.g. long periods of sedentary activity after wake). In addition, to exclude atypical sleep periods, we removed sleep sessions that commenced between 8am and 8pm, sleep sessions that commenced after 8pm but ending on the same day, and split sleep sessions. Importantly, the number of records excluded did not materially differ for records before the lockdown (4.61%) compared to during the lockdown (4.46%). After data filtering, between 766 to 1063 (mean: 898) participants contributed to each timepoint for sleep data (Fig. 1).

**Identification of canonical rest activity patterns**

For each day and from each individual, a rest activity rhythm (RAR) profile was obtained by log-transforming the 24h intraday raw step counts using a natural log function. Each day consisted of 96 15-min interval bins, starting and ending at 00:00 and 23:45 respectively. All valid days across the whole sample from Jan – Apr 2020 (125,851 days) were subsequently fed into a k-means clustering model to identify distinct clusters, or 'basis sets' of daily RAR profiles. This approach enabled quantitative characterization of RAR changes from baseline through lockdown.

The k-means++ algorithm in MATLAB R2016b (Mathworks, Inc., Natick, MA) was used to optimize initialization of the cluster centres[32]. This was done by random selection of the first cluster centre, and subsequently choosing additional cluster centres from the remaining data points with probability proportional to their squared distance from the nearest existing cluster centres.

A range of cluster values was explored (k=2-10) before settling on k=4 for a parsimonious yet meaningful set of clusters. Euclidean distance was used as the distance metric.

**Clustering of persons with similar changes in RAR profile composition during mobility restrictions**

To identify groups of individuals who show similar changes in RAR profile composition with increasing mobility restrictions, we computed proportion of days spent in each RAR profile both before and during the lockdown for each individual. Agglomerative hierarchical clustering was performed using these proportions as feature values on participants with at least 60% of valid days before and during the lockdown (N=670) using Ward's method and the Euclidean distance metric. This clustering method begins by considering each individual as a separate entity, and clustering individuals that are close together in distance. This process was repeated until all individuals were clustered. Inspection of the dendrogram helped identify a 4-cluster solution.

**Statistical analysis**

One-way analyses of variance (ANOVA) and Pearson's chi-squared test of independence were used to analyze sociodemographic differences between the 4 cluster-derived groups. In addition, 2 x 4 mixed ANOVAs were also performed on activity and sleep variables between the different groups with time as the within-subject factor ('Baseline', 'Lockdown') and Group, the between-subject factor. Only subjects who had a minimum of 5 weekdays and 2 weekends in each time point ('Baseline', 'Lockdown') were included in this analysis (N=667). For each variable, data points outside the 1.5 interquartile range were removed from analyses. All statistical analyses were performed in R version 3.6.1.

**REFERENCES**

1. United Nations. 68% of the world population projected to live in urban areas by 2050, says UN. United Nations https://www.un.org/development/desa/en/news/population/2018-revision-of-world-urbanization-prospects.html (2018).

2. Altena, E. *et al.* Dealing with sleep problems during home confinement due to the COVID-19 outbreak: practical recommendations from a task force of the European CBT-I Academy. *J Sleep Res,* e13052 (2020).

3. Wittmann, M., Dinich, J., Merrow, M. & Roenneberg, T. Social jetlag: misalignment of biological and social time. *Chronobiol Int.* **23**, 497-509 (2006).

4. Reutrakul, S. & Van Cauter, E. Interactions between sleep, circadian function, and glucose metabolism: implications for risk and severity of diabetes. *Ann N Y Acad Sci*. **1311**, 151–173 (2014).

5. Cellini, N., Canale, N., Mioni, G. & Costa, S. Changes in sleep pattern, sense of time, and digital media use during COVID-19 lockdown in Italy. *J Sleep Res,* e13074 (2020).

6. Mas, A. & Pallais, A. Valuing alternative work arrangements. *Am Econ Rev.* **107**, 3722-3759 (2017).

7. Sun, S. *et al*. Using smartphones and wearable devices to monitor behavioural changes during COVID-19. Preprint at https://arxiv.org/abs/2004.14331 (2020).

8. Mandolesi, L. *et al.* Effects of physical exercise on cognitive functioning and wellbeing: biological and psychological benefits. *Front Psychol.* **9**, 509 (2018).

9. WHO. Global recommendations on physical activity for health. WHO https://www.who.int/dietphysicalactivity/publications/9789241599979/en/ (2010).

10. Kredlow, M. A., Capozzoli, M. C., Hearon, B. A., Calkins, A. W. & Otto, M. W. The effects of physical activity on sleep: a meta-analytic review. *J Behav Med*. **38**, 427-449 (2015).

11. Mitchell, R. Is physical activity in natural environments better for mental health than physical activity in other environments? *Soc Sci Med.* **91**, 130-134 (2013).

12. Crowley, S. J. & Eastman, C. I. Phase advancing human circadian rhythms with morning bright light, afternoon melatonin, and gradually shifted sleep: can we reduce morning bright-light duration? *Sleep Med.* **16**, 288-297 (2015).

13. Althoff, T. *et al.* Large-scale physical activity data reveal worldwide activity inequality. *Nature* **547**, 336-339 (2017).

14. Audrey, S., Procter, S. & Cooper, A. R. The contribution of walking to work to adult physical activity levels: a cross sectional study. *Int J Behav Nutr Phys Act*. **11**, 37 (2014).

15. Burtscher, J., Burtscher, M. & Millet, G. P. (Indoor) isolation, stress and physical inactivity: vicious circles accelerated by Covid-19? *Scand J Med Sci Sports* https://doi.org/10.1111/sms.13706 (2020).

16. Owen, N., Healy, G., Matthews, C. & Dunstan, D. Too much sitting: the population health science of sedentary nehavior. *Exerc Sport Sci Rev.* **38**, 105-113 (2010).

17. Huang, Y. & Zhao, N. Generalized anxiety disorder, depressive symptoms and sleep quality during COVID-19 outbreak in China: a web-based cross-sectional survey. *Psychiatry Res.* **288**, 112954 (2020).

18. Google. COVID-19 community mobiility reports. Google https://www.google.com/covid19/mobility (2020).

19. Finkelstein, E. A. *et al.* Effectiveness of activity trackers with and without incentives to increase physical activity (TRIPPA): a randomised controlled trial. *Lancet Diabetes Endocrinol.* **4**, 983-995 (2016).

20. Luik, A. I. *et al.* Associations of the 24-h activity rhythm and sleep with cognition: a population-based study of middle-aged and elderly persons. *Sleep Med.* **16**, 850-855 (2015).

21. Zuurbier, L. A. *et al.* Fragmentation and stability of circadian activity rhythms predict mortality: The Rotterdam study. *Am J Epidemiol* **181**, 54-63 (2014).

22. Bloom, N., Liang, J., Roberts, J. & Ying, Z. Does working from home work? Evidence from a Chinese experiment. *Q J Econ.* **130**, 165-218 (2013).

23. Cygan-Rehm, K. & Wunder, C. Do working hours affect health? Evidence from statutory workweek regulations in Germany. *Labour Econ.* **53**, 162-171 (2018).

24. Arnold, J. M., Fitchett, D. H., Howlett, J. G., Lonn, E. M. & Tardif, J.-C. Resting heart rate: a modifiable prognostic indicator of cardiovascular risk and outcomes? *Can J Cardiol* **24**, 3A-8A (2008).

25. Ong, J. L., Tandi, J., Patanaik, A., Lo, J. & Chee, M. Large-scale data from wearables reveal regional disparities in sleep patterns that persist across age and sex. *Sci Rep.* **9**, 3415 (2019).

26. Giuntella, O. & Mazzonna, F. Sunset time and the economic effects of social jetlag: evidence from US time zone borders. *J Health Econ.* **65**, 210-216 (2019).

27. Huang, T. & Redline, S. Cross-sectional and prospective associations of actigraphy-assessed sleep regularity with metabolic abnormalities: the multi-ethnic study of atherosclerosis. *Diabetes Care* **42**, 1422-1429 (2019).

28. Taylor, B. J. *et al.* Bedtime variability and metabolic health in midlife women: the SWAN sleep study. *Sleep* **39**, 457-465 (2016).

29. Vetter, C., Fischer, D., Matera, Joana L. & Roenneberg, T. Aligning work and circadian time in shift workers improves sleep and reduces circadian disruption. *Curr Biol.* **25**, 907-911 (2015).

30. Galea, S., Merchant, R. M. & Lurie, N. The mental health consequences of COVID-19 and physical distancing: the need for prevention and early intervention. *JAMA Intern Med.* https://doi.org/10.1001/jamainternmed.2020.1562 (2020).

31. van Dalfsen, J. H. & Markus, C. R. The influence of sleep on human hypothalamic–pituitary–adrenal (HPA) axis reactivity: a systematic review. *Sleep Med Rev.* **39**, 187-194 (2018).

32. Arthur, D. & Vassilvitskii, S. K-Means++: The Advantages of Careful Seeding. *Proceedings of the Eighteenth Annual ACM-SIAM Symposium on Discrete Algorithms,* 1027-1035 (2007).

**Acknowledgments**: The hiSG study was developed and supported by the Health Promotion Board, Singapore.

**Funding:** Analysis and writing of this sub-study was supported by a grant awarded to MWLC from the National Medical Research Council, Singapore (NMRC/STaR/015/2013).

**Author contributions:** HPB (ZTC, BKLN, DK, WZ and KC) designed the main hiSG study and collected the data. MWLC, JLO and SAAM contributed to this sub-study conceptualization. JLO and TYL analyzed the data. MWLC, JLO, TYL, SAAM and BTTY were responsible for data interpretation. MWLC provided the first draft of the manuscript, while all authors contributed to the final draft.

**Competing interests:** All authors declare no competing interests

**Data and materials availability:** Data cannot be publicly shared due to reasons of confidentiality.  For academic purposes, de-identified participant data can be made available upon reasonable request, and under agreement with the Health Promotion Board, Singapore.

**FIGURES**

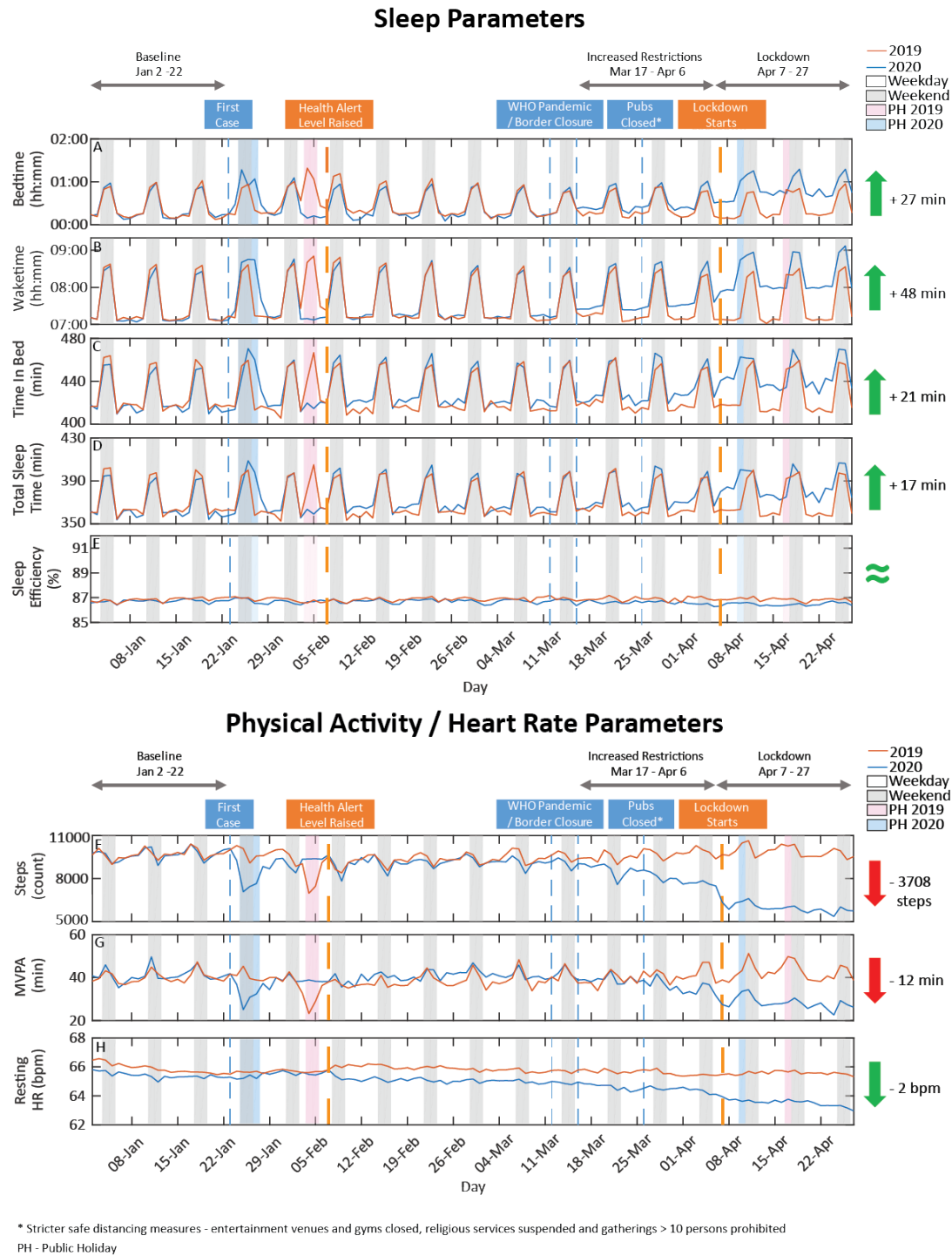

**Fig. 1. Time series plots between Jan 2 - Apr 27, 2020 (blue curves) and Jan 3 - Apr 29, 2019 (red curves) for sleep (top panel) and physical activity/heart rate (bottom panel) parameters.** (A) Bedtime, (B) Waketime, (C) Time In Bed, (D) Total Sleep Time, (E) Sleep Efficiency, (F) Step Counts, (G) Time spent in Moderate-to-Vigorous Physical Activity (MVPA) and (H) Resting Heart Rate. Weekends (gray shaded regions) and public holidays (light blue and pink shaded regions) are also delineated. Dates reflect the 'morning' of each record, such that sleep records always preceded physical activity. Dates in 2019 were shifted by 1 day in order to ensure a matching by day of the week. Key events during the COVID-19 pandemic period ('Baseline', 'Increased Restrictions', and 'Lockdown') are also indicated.

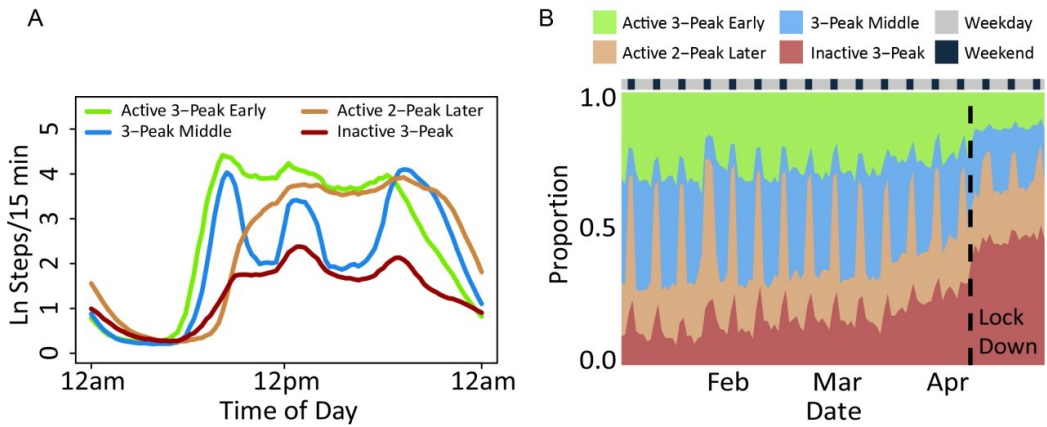

**Fig. 2. RAR profiles.** (A) Centroids for the 4 key RAR profiles determined by k-means clustering of intraday step counts across 125,851 days from Jan-Apr 2020 from all participants for the "Active 3-Peak Early" (green curve, n = 28,723 days), "3-Peak Middle" (blue curve, n = 32,981 days), "Active 2-Peak Later" (brown curve, n = 30,635 days) and "Inactive 3-Peak" (red curve, n = 33,512 days) clusters. Clusters were differentiated by timing of morning rise and evening drop as well as magnitude of steps across the 24h day. (B) Proportion of time spent in the 4 RAR profiles from Jan-Apr 2020. Before the lockdown, weekday rhythms primarily consisted of the "Active 3-Peak Early" and "3-Peak Middle" profiles, while weekend rhythms mainly consisted of the "Active 2-Peak Later" and "Inactive 3-Peak" profiles. After the lockdown, a clear increase in the proportion of time spent in the "Active2-Peak Later" and "Inactive 3-Peak" profiles was observed, together with an attenuation of weekday-weekend rhythms.

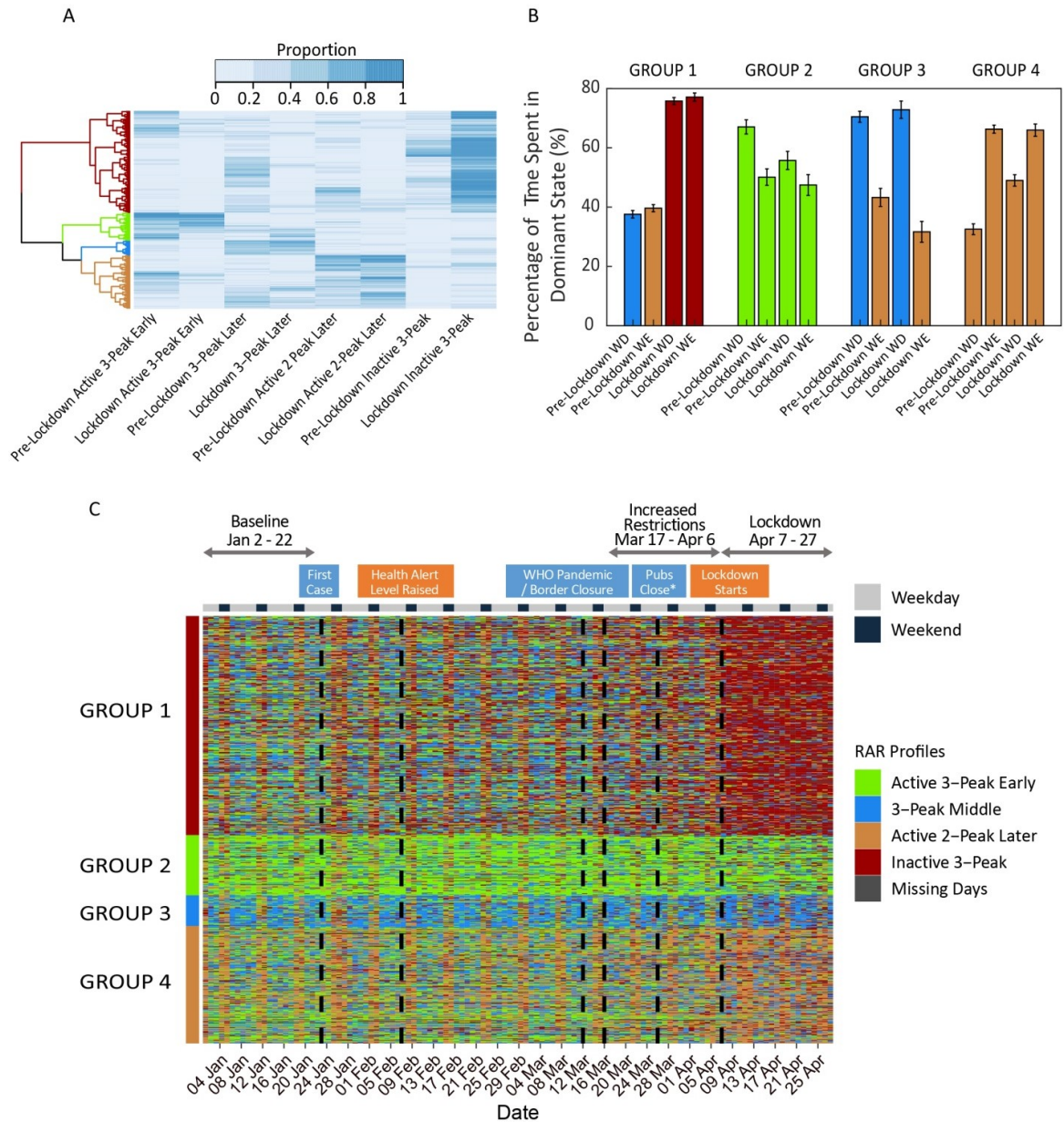

**Fig. 3. Identification of groups with similar RAR changes before and during the lockdown.** (A) Hierarchical clustering of participants based on proportion of time spent in the 4 RAR profiles before and after the lockdown. Visual inspection of the dendrogram identified 4 groups of participants with similar changes in the patterns of RAR profiles before and during the lockdown. (B) Proportion of time spent in the dominant RAR profiles on weekdays and weekends for each group, before and after the lockdown. (C) RAR profiles across days (columns) by participant (rows) ordered by groups partitioned from the hierarchical clustering. Groups are coloured by their dominant RAR profile colour for ease of reference.

**TABLES**

**Table 1. Sociodemographic characteristics of hiSG study participants**

| Characteristics | Statistics |
|---|---|
| Age in years, mean (SD), range | 30·94 (4·62), 21-41 |
| Gender, N (%) | |
|     Female | 941 (51·64) |
| Ethnicity, N (%) | |
|     Chinese | 1718 (94·19) |
|     Malay | 31 (1·70) |
|     Indian | 48 (2·63) |
|     Others | 27 (1·48) |
| Family Status, N (%) | |
|     Single | 934 (51·21) |
|     Married with No Children | 376 (20·61) |
|     Married with Children | 484 (26·54) |
|     Separated/Divorced/Widowed | 30 (1·64) |
| Household Income in SGD, N (%) | |
|     < $2k | 180 (9·87) |
|     $2k - $3·9k | 721 (39·53) |
|     $4k- $5·9k | 511 (28·02) |
|     $6k - $7·9k | 172 (9·43) |
|     $8k - $9·9k | 96 (5·26) |
|     ≥ $10k | 144 (7·89) |
| Highest Education, N (%) | |
|     Degree | 1570 (86·07) |
|     No Degree | 254 (13·93) |

**Table 2. Means of Sleep and Physical Activity Variables in 2019 and 2020.**

Sleep variables in 2020 by phase

|  | Jan 2-22, 2020 (Baseline) | | Mar 17-Apr 6, 2020 (Increased Restrictions) | | Apr 7-27, 2020 (Lockdown) | |
| --- | --- | --- | --- | --- | --- | --- |
|  | WD | WE | WD | WE | WD | WE |
| **Bedtime (hh:mm)** | 00:13 | 00:53 | 00:24 | 00:56 | 00:43 | 01:12 |
| **Waketime (hh:mm)** | 07:07 | 08:26 | 07:29 | 08:36 | 08:02 | 08:57 |
| **Time In Bed, TIB (h)** | 6·91 | 7·54 | 7·08 | 7·66 | 7·31 | 7·75 |
| **Total Sleep Time, TST (h)** | 5·99 | 6·54 | 6·12 | 6·64 | 6·32 | 6·70 |
| **Wake After Sleep Onset, WASO (h)** | 0·92 | 1·00 | 0·95 | 1·03 | 0·99 | 1·04 |
| **Sleep Efficiency (%)** | 86·7 | 86·8 | 86·6 | 86·6 | 86·4 | 86·6 |

Physical activity variables in 2020 by phase

|  | Jan 2-22, 2020 (Baseline) | | Mar 17-Apr 6, 2020 (Increased Restrictions) | | Apr 7-27, 2020 (Lockdown) | |
| --- | --- | --- | --- | --- | --- | --- |
|  | WD | WE | WD | WE | WD | WE |
| **Steps (count)** | 9729 | 9484 | 8321 | 7942 | 5905 | 6065 |
| **Moderate to Vigorous Physical Activity (min)** | 39·8 | 43·9 | 36·4 | 38·9 | 27·4 | 29·9 |
| **Resting Heart Rate (bpm)** | 65·4 | 65·3 | 64·5 | 64·4 | 63·5 | 63·5 |

Control comparison of sleep variables in Jan 2019 and Jan 2020

|  | Jan 3-23, 2019 | | Jan 2-22, 2020 | |
| --- | --- | --- | --- | --- |
|  | WD | WE | WD | WE |
| **Bedtime (hh:mm)** | 00:12 | 00:54 | 00:13 | 00:53 |
| **Waketime (hh:mm)** | 07:09 | 08:33 | 07:30 | 08:26 |
| **Time In Bed, TIB (h)** | 6·94 | 7·65 | 6·91 | 7·54 |
| **Total Sleep Time, TST (h)** | 6·02 | 6·64 | 5·99 | 6·54 |
| **Wake After Sleep Onset, WASO (h)** | 0·92 | 1·00 | 0·92 | 1·00 |
| **Sleep Efficiency (%)** | 86·8 | 86·9 | 86·7 | 86·8 |

Control comparison of physical activity variables in Jan 2019 and Jan 2020

|  | Jan 3-23, 2019 | | Jan 2-22, 2020 | |
| --- | --- | --- | --- | --- |
|  | WD | WE | WD | WE |
| **Steps (count)** | 9769 | 9513 | 9729 | 9484 |
| **Moderate to Vigorous Physical Activity (min)** | 38·8 | 43·3 | 39·8 | 43.9 |
| **Resting Heart Rate (bpm)** | 65·9 | 65·9 | 65·4 | 65·3 |

**Table 3. Socio-demographic, daily activity and daily sleep measures by group.**

|  | Group 1 (M±SD) n = 301-344 | Group 2 (M±SD) n = 78-94 | Group 3 (M±SD) n = 39-50 | Group 4 (M±SD) n = 158-182 | p |
|---|---|---|---|---|---|
| **Socio-demographics** | | | | | |
| Age (y) | 31·25 (4·46)[24] | 32·87 (4·23)[1] | 32·52 (4·05) | 32·78 (4·23)[1] | < 0·0001 |
| Sex-Females (%) | 50·00 | 46·02 | 52·00 | 47·25 | 0·87 |
| Ethnicity (%) | | | | | 0·20 |
|   Chinese | 94·48 | 93·62 | 98·00 | 91·76 | |
|   Malay | 2·33 | 3·19 | 0·00 | 0·55 | |
|   Indian | 2·04 | 1·06 | 2·00 | 3·85 | |
|   Others | 1·16 | 2·13 | 0·00 | 3·85 | |
| Education-Degree holders (%) | 84·88 | 84·04 | 100·00 | 83·52 | 0·03 |
| Household Income (%) | | | | | 0·06 |
|   < $2k | 10·17 | 11·70 | 2·00 | 10·99 | |
|   $2k - $3·9k | 40·70 | 44·68 | 26·00 | 32·42 | |
|   $4k - $5·9k | 26·16 | 13·83 | 34·00 | 28·57 | |
|   $6k - $7·9k | 9·01 | 13·83 | 16·00 | 12·09 | |
|   $8k - $9·9k | 6·11 | 9·57 | 6·00 | 6·60 | |
|   ≥ $10k | 7·85 | 6·38 | 16·00 | 9·34 | |
| Family Status (%) | | | | | < 0·0001 |
|   Single | 57·85[a] | 35·11 | 40·00 | 35·17[b] | |
|   Married with no children | 20·64 | 7·45 | 28·00 | 18·68 | |
|   Married with children | 20·06[b] | 52·13[a] | 30·00 | 43·96[a] | |
|   Separated/divorced/widowed | 1·45 | 5·32[b] | 2·00 | 2·20 | |
| **Health** | | | | | |
| BMI (kg/m$^2$) | 23·25 (4·12) | 23·40 (3·88) | 21·85 (2·25)[4] | 23·61 (3·83)[3] | 0·04 |

|  | Baseline | Lockdown | Baseline | Lockdown | Baseline | Lockdown | Baseline | Lockdown | |
|---|---|---|---|---|---|---|---|---|---|
| **Daily Activity** | | | | | | | | | |
| Total Steps | 9516·81 (2637·43) | 4683·34 (2381·67)*** | 12010·51 (3055·46) | 9268·30 (3374·22)*** | 10959·78 (2567·95) | 8861·49 (3302·52)*** | 11275·40 (2780·22) | 8309·44 (2934·34)*** | < 0·0001 < 0·0001 |

| Variable | | | | | | | | | P |
|---|---|---|---|---|---|---|---|---|---|
| | | | | | | | | | < 0·0001 |
| MVPA (min) | 41·27 (24·25) | 22·64 (18·33)*** | 52·22 (29·77) | 39·78 (24·13)*** | 47·74 (25·96) | 45·06 (28·47) | 50·94 (29·04) | 36·69 (23·16)*** | < 0·0001 < 0·0001 0·0001 |
| Resting Heart Rate (bpm) | 64·24 (6·51) | 63·04 (6·65)*** | 63·76 (6·51) | 62·22 (6·37)*** | 63·28 (6·48) | 61·57 (6·58)*** | 63·56 (6·23)*** | 62·11 (6·49)*** | 0·38 < 0·0001 0·57 |
| **Daily Sleep** | | | | | | | | | |
| Bedtime (hh:mm) | 00:25 (55·50) | 00:55 (68·76)*** | 11:43 (46·68) | 11:48 (44·50) | 00:00 (37·00) | 00:08 (40·79) | 00:25 (50·61) | 00:57 (57·62)*** | < 0·0001 < 0·0001 < 0·0001 |
| Wake Time (hh:mm) | 07:31 (54·38) | 08:24 (64·52)*** | 06:39 (37·19) | 06:54 (36·79)** | 06:58 (25·46) | 07:25 (32·71)*** | 07:38 (48·69) | 08:28 (49·93)*** | < 0·0001 < 0·0001 < 0·0001 |
| Time In Bed (h) | 7·07 (38·64) | 7·44 (43·05)*** | 6·97 40·68 | 7·16 (40·27)* | 6·96 (30·25) | 7·28 (34·91)*** | 7·15 (42·10) | 7·48 (45·28)*** | 0·01 < 0·0001 0·11 |
| Total Sleep Time (h) | 6·13 (32·92) | 6·43 (37·49)*** | 6·05 (38·02) | 6·19 (36·37) | 6·05 (26·58) | 6·30 (30·05)** | 6·19 (36·06) | 6·46 (39·86)*** | 0·03 < 0·0001 0·14 |

[1] Significantly different from Group 1
[2] Significantly different from Group 2
[3] Significantly different from Group 3
[4] Significantly different from Group 4
[a] Cell's observed proportion significantly higher than its expected proportion
[b] Cell's observed proportion significantly lower than its expected proportion

P values under P column for Daily Activity and Daily Sleep are ordered by Main Effect of Group, Main Effect of Time, Interaction Effect of Group and Time
Significance values under Daily Activity and Daily Sleep refers to a significant difference from baseline

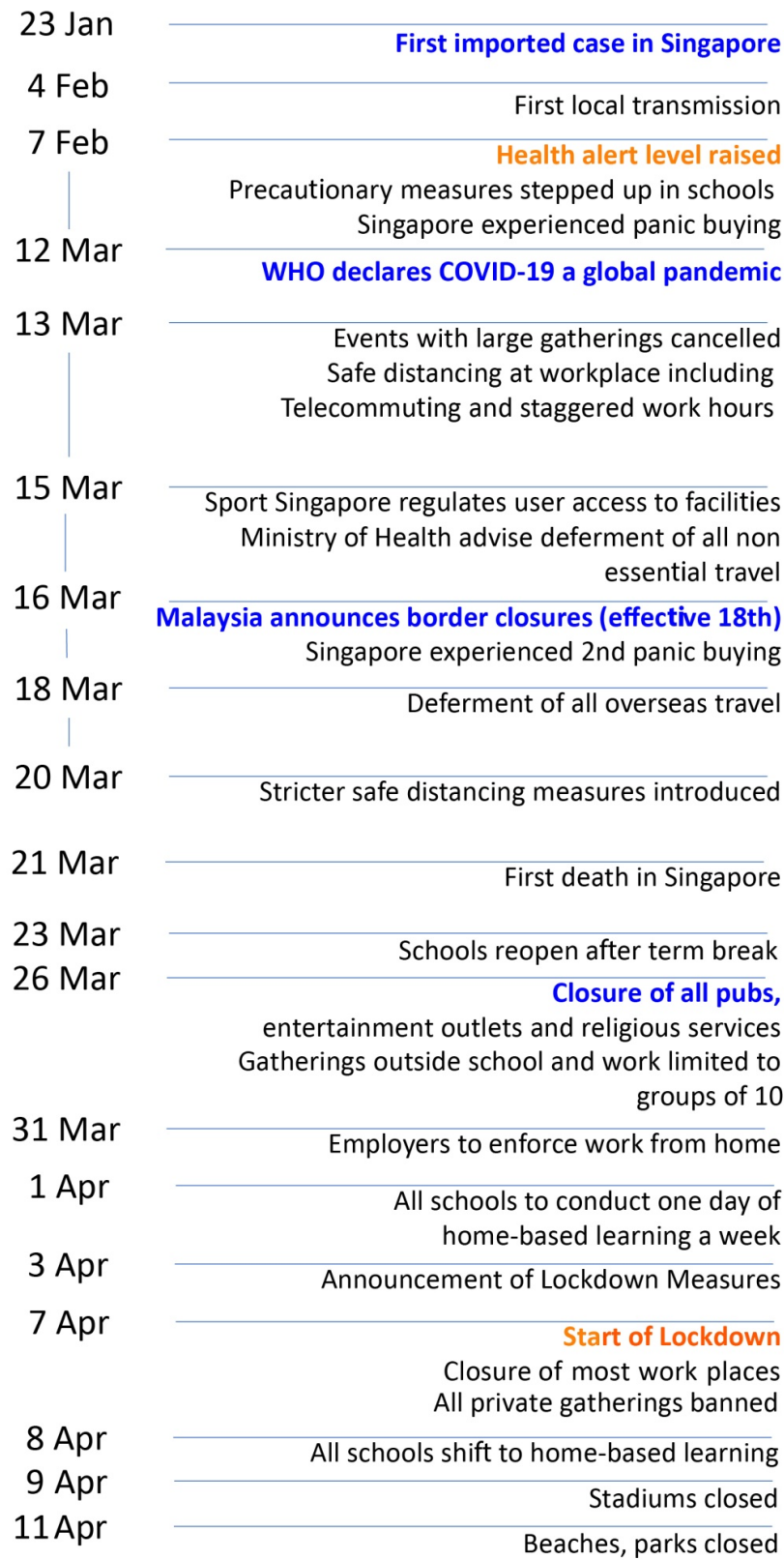

**Fig. S1. Timeline of key COVID events in Singapore.**

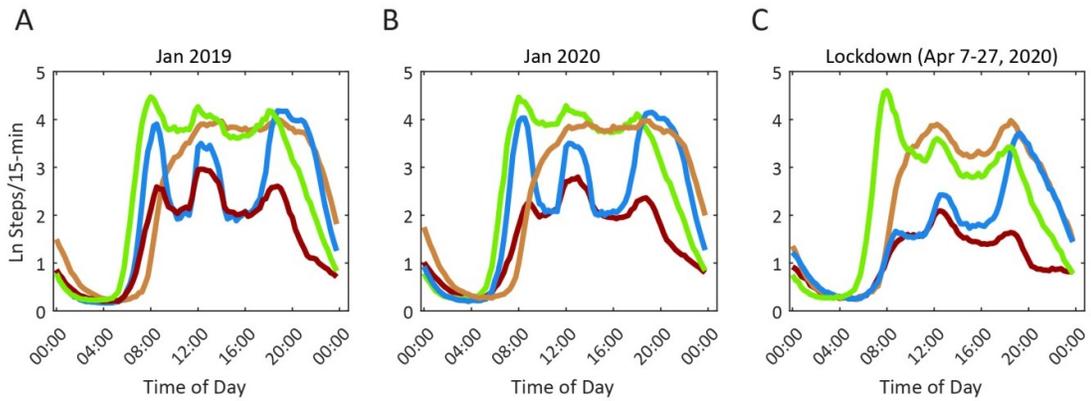

**Fig. S2. k-means clustering of RAR profiles in Jan 2019, Jan 2020 and during the lockdown.** RAR profiles were highly similar between Jan 2019 and Jan 2020 (r > 0.99). After the lockdown RAR profiles were slightly attenuated, but still highly similar to the profiles estimated using the full dataset from Jan-Apr 2020 (r: 0.84-0.98).

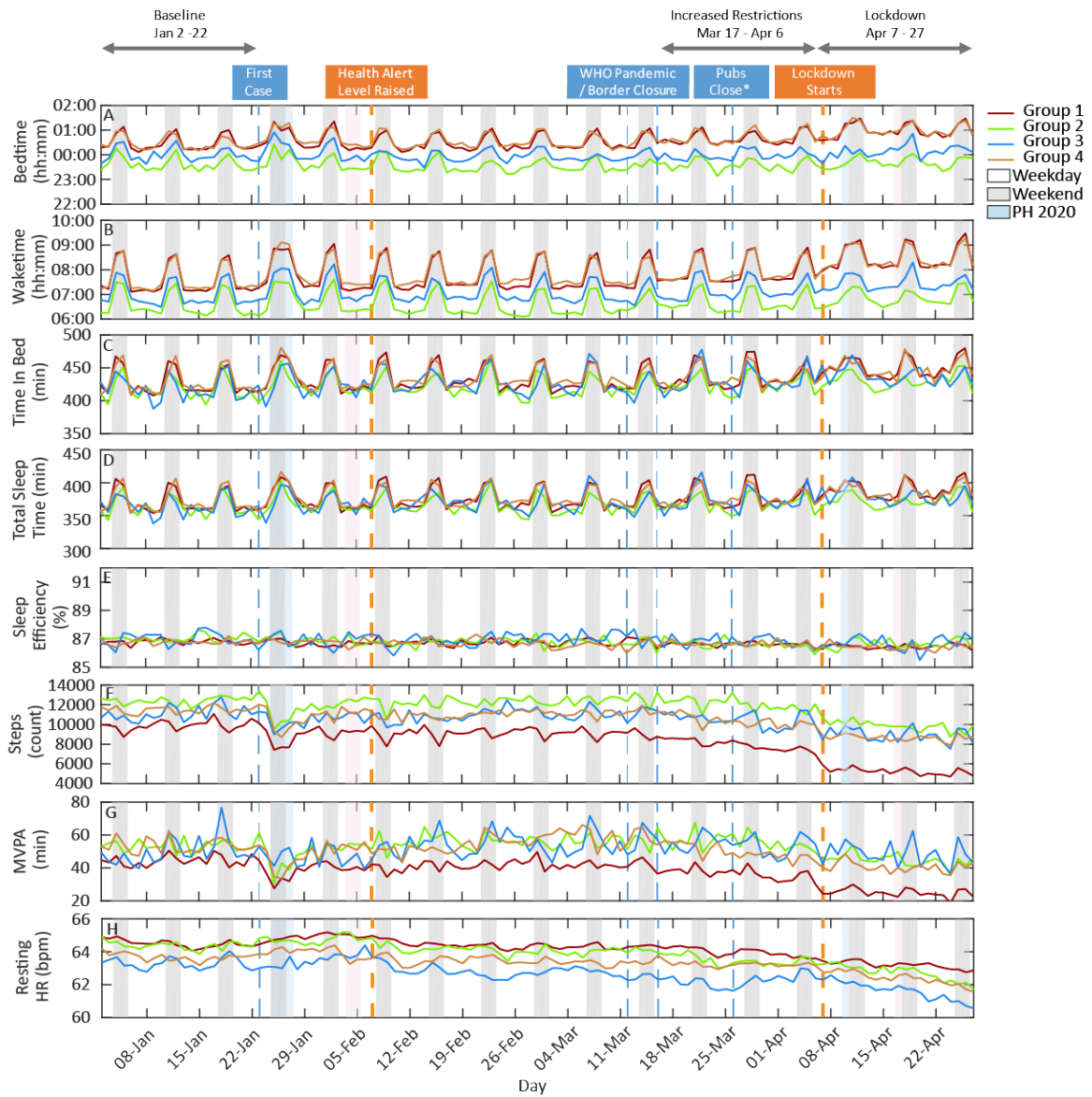

**Fig. S3. Time series plots between Jan 2 - Apr 27, 2020 (blue curves) and Jan 3 - Apr 29, 2019 (red curves).** (A) Bedtime, (B) Waketime, (C) Time In Bed, (D) Total Sleep Time, (E) Sleep Efficiency, (F) Step Counts, (G) Time spent in Moderate-to-Vigorous Physical Activity (MVPA) and (H) Resting Heart Rate for the 4 groups identified by the hierarchical clustering. Weekends (gray shaded regions) and public holidays (light blue and pink shaded regions) are also delineated. Dates reflect the 'morning' of each record, such that sleep records always preceded physical activity. Key events during the COVID-19 pandemic period ('Baseline', 'Increased Restrictions', and 'Lockdown') are also indicated.

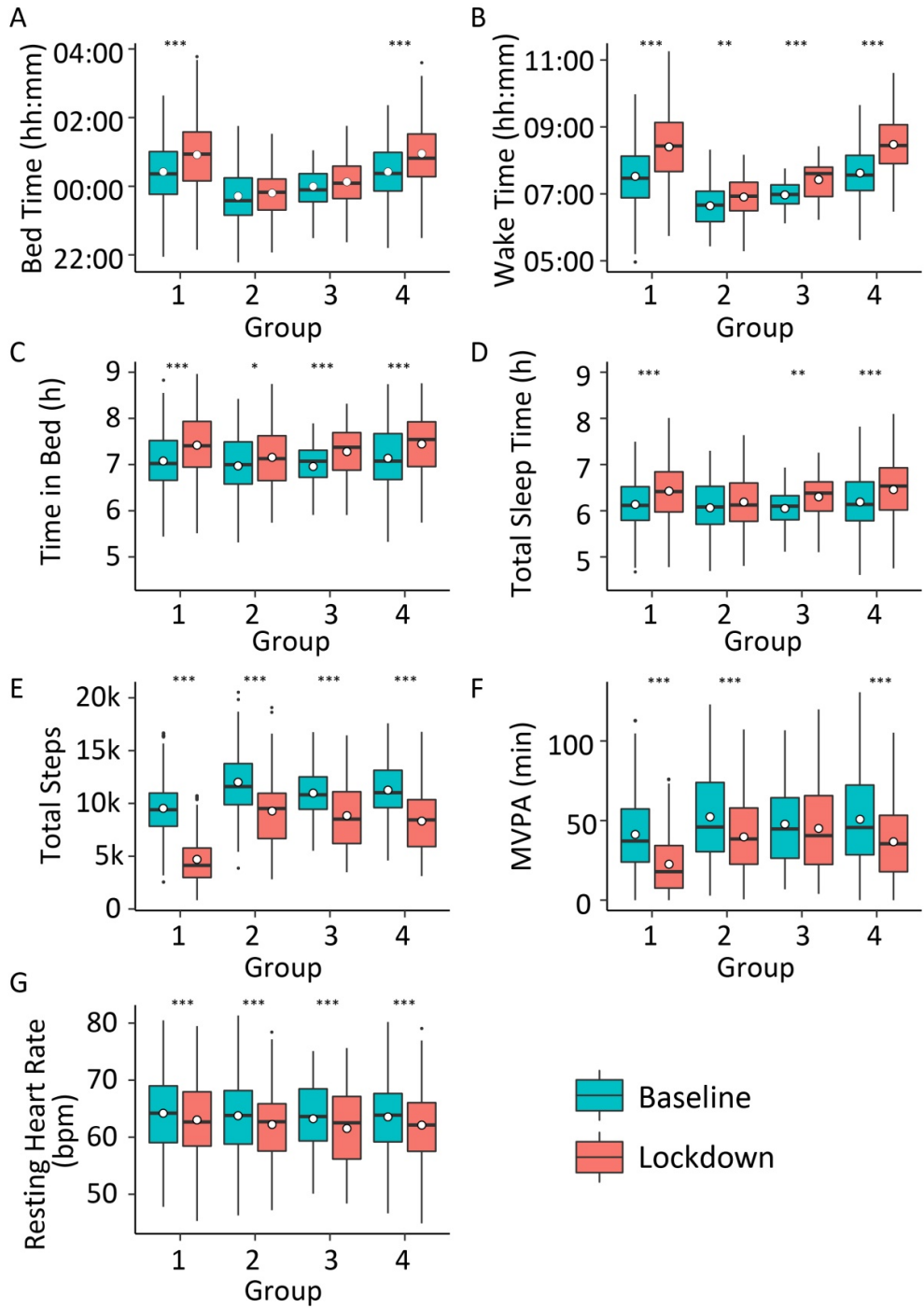

**Fig. S4. Boxplots for comparisons between groups identified by the hierarchical clustering, during the 'Baseline' and 'Lockdown' period**. (A) Bedtime, (B) Waketime, (C) Time In Bed, (D) Total Sleep Time, (E) Step Counts, (F) Time spent in Moderate-to-Vigorous Physical Activity (MVPA) and (G) Resting Heart Rate. Asterisks denote significant pairwise comparisons between 'Baseline' and 'Lockdown' periods for each group. *p<0.05, **p<0.01, ***p<0.001.